\PassOptionsToPackage{unicode}{hyperref}
\PassOptionsToPackage{hyphens}{url}
\documentclass[11pt,]{article}
\usepackage{authblk}
\usepackage{lmodern}
\usepackage{amsmath}
\usepackage{amssymb}
\usepackage{bbm}
\usepackage{dsfont}
\DeclareMathOperator*{\argmax}{arg\,max}
\usepackage[utf8]{inputenc}
\usepackage{moreverb}
\usepackage{graphicx} 
\usepackage{mwe} 
\usepackage[FIGTOPCAP]{subfigure}

\usepackage{setspace} 
\usepackage[francais]{babel} 

\usepackage{hyperref}

\usepackage{ifxetex,ifluatex}
\ifnum 0\ifxetex 1\fi\ifluatex 1\fi=0 
  \usepackage[T1]{fontenc}
  \usepackage[utf8]{inputenc}
  \usepackage{textcomp} 
  \usepackage{amssymb}
\else 
  \usepackage{unicode-math}
  \defaultfontfeatures{Scale=MatchLowercase}
  \defaultfontfeatures[\rmfamily]{Ligatures=TeX,Scale=1}
\fi
\IfFileExists{upquote.sty}{\usepackage{upquote}}{}
\IfFileExists{microtype.sty}{
  \usepackage[]{microtype}
  \UseMicrotypeSet[protrusion]{basicmath} 
}{}
\makeatletter
\@ifundefined{KOMAClassName}{
  \IfFileExists{parskip.sty}{%
    \usepackage{parskip}
  }{
    \setlength{\parindent}{0pt}
    \setlength{\parskip}{6pt plus 2pt minus 1pt}}
}{
  \KOMAoptions{parskip=half}}
\makeatother
\usepackage{xcolor}
\IfFileExists{xurl.sty}{\usepackage{xurl}}{} 
\IfFileExists{bookmark.sty}{\usepackage{bookmark}}{\usepackage{hyperref}}
\usepackage[margin=1in]{geometry}
\usepackage{color}
\usepackage{fancyvrb}

\DefineVerbatimEnvironment{Highlighting}{Verbatim}{commandchars=\\\{\}}
\usepackage{framed}
\definecolor{shadecolor}{RGB}{248,248,248}

\newcommand{\rev}[1]{{#1}}
\newcommand{\bfs}[1]{\boldsymbol{#1}}

\usepackage{graphicx}
\makeatletter
\def\maxwidth{\ifdim\Gin@nat@width>\linewidth\linewidth\else\Gin@nat@width\fi}
\def\maxheight{\ifdim\Gin@nat@height>\textheight\textheight\else\Gin@nat@height\fi}
\makeatother
\setkeys{Gin}{width=\maxwidth,height=\maxheight,keepaspectratio}
\makeatletter
\def\fps@figure{htbp}
\makeatother
\ifluatex
  \usepackage{selnolig}  
\fi

\title{Describing complex disease
progression using joint latent class models for multivariate
longitudinal markers and clinical endpoints}

\author{Cécile Proust-Lima$^{1,2,*}$, Tiphaine Saulnier$^{1}$, Viviane Philipps$^{1}$,Anne Pavy-Le Traon$^{3}$, \\
Patrice Péran$^{4}$, Olivier Rascol$^{3}$, Wassilios G. Meissner$^{5,6}$, Alexandra Foubert-Samier$^{1,2,5}$}
\date{}

\begin{document}
\maketitle

\textit{$^{1}$ Univ. Bordeaux, Inserm, BPH, U1219, F-33000 Bordeaux, France}\\
\textit{$^{2}$ Inserm, CIC1401-EC, F-33000 Bordeaux, France}\\
\textit{$^{3}$ MSA Reference Center and CIC-1436, Department of Clinical Pharmacology and Neurosciences, NeuroToul COEN Center, University of Toulouse 3, CHU of Toulouse, INSERM, Toulouse, France}\\
\textit{$^{4}$ ToNIC, Toulouse NeuroImaging Center, Univ Toulouse, Inserm, UPS, F-31000 Toulouse, France}\\
\textit{$^{5}$ Univ. Bordeaux, CNRS, IMN, UMR5293, F-33000 Bordeaux, France}\\
\textit{$^{6}$ Dept. Medicine, University of Otago, Christchurch, and New Zealand Brain Research Institute, Christchurch, New Zealand}\\
\textit{$^{*}$ Correspondence: C\'ecile Proust-Lima, Inserm UMR1219, 146 rue L\'eo Saignat, 33076 Bordeaux Cedex, France. cecile.proust-lima@inserm.fr}

\newpage
\textbf{Abstract:}\\
Neurodegenerative diseases are characterized by numerous markers of progression and clinical endpoints. For instance, Multiple System Atrophy (MSA), a rare neurodegenerative synucleinopathy, is characterized by various combinations of progressive autonomic failure and motor dysfunction, and a very poor prognosis.
Describing the progression of such complex and multi-dimensional diseases is particularly difficult. One has to simultaneously account for the assessment of multivariate markers over time, the occurrence of clinical endpoints, and a highly suspected heterogeneity between patients. Yet, such description is crucial for understanding the natural history of the disease, staging patients diagnosed with the disease, unravelling subphenotypes, and predicting the prognosis.
Through the example of MSA progression, we show how a latent class approach \rev{modeling multiple repeated markers and clinical endpoints} can help describe complex disease progression and identify subphenotypes for exploring new pathological hypotheses. 
The \rev{proposed} joint latent class model includes class-specific multivariate mixed models to handle multivariate repeated biomarkers possibly summarized into latent dimensions and class-and-cause-specific proportional hazard models to handle time-to-event data. Maximum likelihood estimation procedure, \rev{validated through simulations} is available in the lcmm R package. 
In the French MSA cohort comprising data of 598 patients during up to 13 years, five subphenotypes of MSA were identified that differ by the sequence and shape of biomarkers degradation, and the associated risk of death. In posterior analyses, the five subphenotypes were used to explore the association between clinical progression and external imaging and fluid biomarkers, while properly accounting for the uncertainty in the subphenotypes membership. 


\maketitle


\newpage
\section{Introduction} \label{intro}

Some diseases are characterized by numerous markers of progression. Although not specific to, this is particularly the case in neurodegenerative diseases where pathological brain changes may induce multiple clinical signs on which the progression of a patient is assessed. Alzheimer’s disease involves the progressive impairment over decades of cerebral regions, multiple cognitive functions, functional dependency, and even depressive symptomatology or anxiety \cite{amieva2008,jack2013}. Parkinson disease is a polymorph disease including progressive motor impairment, cognitive and behavioural disorders, and autonomic failure \cite{weintraub2018,swick2014}. Multiple System Atrophy (MSA), a rare neurodegenerative synucleinopathy with annual incidence of 3/100,000 individuals \cite{wenning2004MSA}, is also characterized by the combinations of multiple dimensions, including autonomic failure, parkinsonism and cerebellar ataxia \cite{meissner2019}. 

Describing the progression of such complex and multi-dimensional diseases is particularly difficult. One has to simultaneously account for the assessment of multivariate markers over time, the occurrence of clinical endpoints (e.g., death, extreme dependency), and the suspected heterogeneity between patients. Yet, such description is crucial for understanding the natural history of the disease, staging patients diagnosed with the disease, unravelling subphenotypes, \rev{identifying novel therapeutic targets} and predicting the prognosis. \\

When interested in the change over time of markers along with occurrence of endpoints, the dedicated statistical approach is the joint modelling methodology which simultaneously models the trajectory over time of a marker and the risk of an event when those two are correlated \cite{henderson2000,rizopoulos2012,proust-lima2014}. Traditionally based on the so called «shared random effect» paradigm, joint models usually focus on how longitudinal markers impact the risk of an event by including some predictor of the marker trajectory in the time-to-event model. This is particularly useful to quantify the association between pre-determined characteristics of an endogenous marker and the clinical endpoints \cite{rizopoulos2012, sene2014}. However, it might not be the best way to explore in a hollistic way a complex disease progression which involves multiple markers along with clinical endpoints. In that perspective, joint latent class models (JLCMs) \cite{lin2002,proust-lima2014,liu2020,huang2018}, another family of joint models, constitute a relevant alternative. JLCMs assume that the population of patients is heterogeneous, and that this heterogeneity explains why patients experiment different marker profiles and different event risks. This paradigm is much more descriptive than traditional joint models but apprehends the suspected heterogeneity present in many contexts and does not assume any particular nature of association between the marker and the event \cite{proust-lima2014}. Over the years, several extensions of JLCM were proposed regarding the nature of the survival data with competing events \cite{proust-lima2016}, recurrent events \cite{han2007} or event history \cite{rouanet2016}, or regarding the nature of the longitudinal data by considering several markers measuring the same underlying phenomenon \cite{proust-lima2009,proust-lima2016}, multiple Gaussian markers from high dimensional gene expression \rev{with a regularization step} \cite{sun2019} and \rev{multiple Gaussian markers} subject to limits of detection \cite{li2020}. \\

\rev{In this work, we aimed to leverage the latent class approach to analyze multiple repeated progression markers over time and clinical endpoints, with the final goal of retrieving the subphenotypes of MSA progression and linking them with external biomarker information. Our contribution is fourfold. First, we developed a full methodology for the estimation of joint latent class models for multidimensional longitudinal data and survival time (possibly with competing causes). This model extends beyond the literature by handling multidimensional longitudinal data when Proust-Lima et al. \cite{proust-lima2016} considered multivariate longitudinal markers regrouped into a uni-dimensional latent process, and by considering multivariate Gaussian and non-Gaussian continuous markers possibly regrouped into distinct latent dimensions along with multi-cause (left-truncated) time-to-event when Sun et al \cite{sun2019} and Li et al \cite{li2020} considered Gaussian markers possibly subject to detection limit, and classical survival data. Second, our methodology is made available to the community with a dedicated function in the user-friendly lcmm R software for latent process and latent class models estimation \cite{proust-lima2017} along with documentation. Third, one critical but often ignored aspect of latent class models is the interpretation of the final latent classes, and their association with external information (covariates or outcomes). The uncertainty and miss-classification of any posterior class assignment has to be carefully accounted for to avoid spurious associations \cite{elliott2020, clark2009,bakk2014}. Following previous works in non-longitudinal mixture modeling \cite{bakk2018}, our method includes two-stage posterior regressions for linking the latent class structure with external information while properly accounting for the uncertainty of the latent class structure. Fourth, we extensively describe a case study in MSA progression to show step-by-step how the JLCM methodology can help describe complex disease progression, identify disease subphenotypes and explore new research hypotheses.\\
}

\rev{Next section introduces the motivating MSA data. Section 3 details the multivariate JLCM methodology including the model, the maximum likelihood estimation procedure and the strategy to associate the latent class structure with external information. Section 4 assesses the finite sample performances through simulations. Section 5 is dedicated to the MSA application. Finally, Section 6 concludes. \\
}

\section{The French Multi-System Atrophy Cohort} \label{MSA}

The French MSA cohort has been created in 2007 at the two sites (Toulouse and Bordeaux university hospitals) of the French Reference Center for MSA. After inclusion, patients are usually seen at least once a year by a movement disorder specialist with a clinical assessment that includes demographic information, medical history, neurological examination, diagnostic certainty and subtype, and a standardized clinical evaluation using the Unified MSA Rating Scale (UMSARS) \cite{wenning2004}. There is also a continuous search for death occurrence with a reporting of the exact date of death along with the cause of death. \\

The first objective of our study was to describe the clinical progression of MSA, and to uncover potential heterogeneous disease phenotypes using latent classes. We focused on the repeated measures of six UMSARS-derived markers regrouped in three different dimensions:
\begin{itemize}
\item[-] Motor and function dimension assessed by a subscore of Activities of Daily Living (UMSARS I) and a subscore of Motor examination (UMSARS II);
\item[-] Supine blood pressure (BP) assessed by the systolic BP and the diastolic BP. 
\item[-] The orthostatic change in BP assessed by the maximum change in systolic BP and in diastolic BP between supine position and standing position during 10 minutes (UMSARS III). 
\end{itemize}
We leveraged the data of the 598 patients enrolled between 2007 and 2019 who had at least one measure of each marker during the follow-up and no missing information on the main known MSA characteristics: gender, age at inclusion, duration since first symptoms, subtype of MSA (Cerebellar or Parkinson), level of diagnosis certainty (possible or probable). See Table \ref{TabDescr} for the sample description. Patients entered the study on average 4.5 years after the first symptoms onset (min-max=0-24 years). During follow-up, 309 patients died with a median survival since first symptoms of 6.65 years (95\% confidence interval [6.18,7.13] years).

\begin{figure}[!p]
\centering\includegraphics[width=0.9\textwidth]{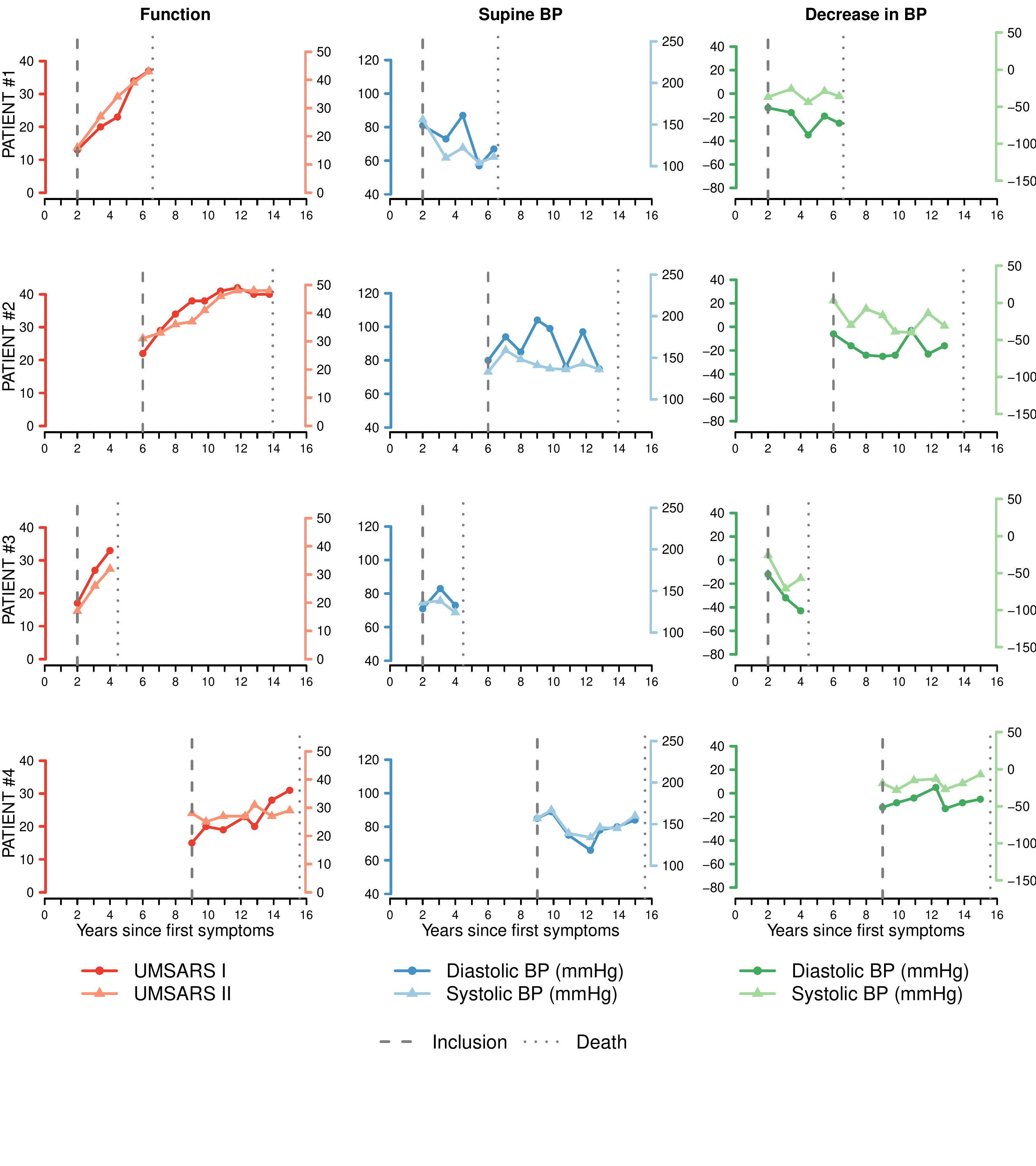}
\caption{Observed marker trajectories for 4 randomly selected deceased patients according to the time since first symptoms. Are reported the time of inclusion in the cohort and the time of death. The 6 markers under study are the UMSARS I and II subscores both measuring functional dependency, the systolic and diastolic supine blood pressure (BP), and the maximum decrease in systolic and diastolic BP when standing up.}
\label{FigSpagh}
\end{figure}

To describe the natural history of MSA, we considered the time since first symptoms as a proxy of the time since disease onset. Figure \ref{FigSpagh} describes the individual observed trajectories for the 6 markers of progression under study of 4 randomly selected patients according to the time since first symptoms. \\

\begin{table}[!p]
\begin{center}
\begin{tabular}{lcc}
\hline
Characteristic & N (\%) & mean $\pm$ sd \\
            \hline
            \underline{At baseline} \\
                \hspace{0.5cm} Sex \\
                    \hspace{1cm} Male & 297 (49.67\%) & \\
                    \hspace{1cm} Female & 301 (50.33\%) & \\
                \hspace{0.5cm} Hospital \\
                    \hspace{1cm} Bordeaux & 309 (51.67\%) & \\
                    \hspace{1cm} Toulouse & 289 (48.33\%) & \\
                \hspace{0.5cm} Type of diagnosis \\
                    \hspace{1cm} MSA-C, with predominant cerebellar impairment & 198 (33.11\%) & \\
                    \hspace{1cm} MSA-P, with predominant parkinsonism & 400 (66.89\%) & \\
                \hspace{0.5cm} Diagnosis certainty \\
                    \hspace{1cm} Possible & 144 (24.08\%) & \\
                    \hspace{1cm} Probable & 454 (75.92\%) & \\
                \hspace{0.5cm} Age at inclusion & & 65.05 $\pm$ 8.09 \\
                \hspace{0.5cm} Years since first symptoms & & 4.54 $\pm$ 2.58 \\
                \hspace{0.5cm} Clinical markers & &  \\
                    \hspace{1cm} total UMSARS-I score & & 20.88 $\pm$ 7.35 \\
                    \hspace{1cm} total UMSARS-II score & & 23.11 $\pm$ 8.25 \\
                    \hspace{1cm} supine diastolic BP (in mmHg) & & 81.84 $\pm$ 13.37 \\
                    \hspace{1cm} supine systolic BP (in mmHg) & & 140.8 $\pm$ 23.52 \\
                    \hspace{1cm} maximum drop of diastolic BP (in mmHg) & & -17.37 $\pm$ 14.88 \\
                    \hspace{1cm} maximum drop of systolic BP (in mmHg) & & -33.87 $\pm$ 25.16 \\
            \hline
            \underline{During follow-up} \\
                \hspace{0.5cm} Repeated visits per patient & & 2.98 $\pm$ 2.08 \\
                \hspace{0.5cm} Length of follow-up (years) & & 6.96 $\pm$ 3.33 \\
                \hspace{0.5cm} Death & 309 (51.67\%) & \\
            \hline
\end{tabular}
\end{center}
\caption{Description at baseline and over follow-up of the 598 MSA patients under study in the French MSA cohort. }\label{TabDescr}
\end{table}

The second objective of this application was to explore to what extent these phenotypes were associated with biological biomarkers which constitute potential therapeutic targets. Indeed, additional assessments were undertaken on small subsamples of the cohort to explore new research hypotheses. A MRI-subsample of 86 patients underwent a T1-weighted volumetric brain Magnetic Resonance Imaging (MRI) with a focus on the volumes of regions particularly involved in the neurodegenerative process such as the cerebellum (gray and white matter), the putamen and the pons. These volumes were measured using the FreeSurfer's image analysis pipeline (version 6). Another subsample underwent further cerebro-spinal fluid (CSF) and serum measurements of total $\alpha$-synuclein concentration (for 23 patients) and neurofilament light chain (NfL) (for 52 patients). See Table S29 in supplementary materials for a description of the three subsamples.

\section{Methodology} \label{method}

\subsection{The joint latent class model} \label{JLCM}

\subsubsection{Latent class structure}
The latent class methodology relies on a latent class/group structure underlying the variables under study. Let us consider a sample of $N$ subjects ($i=1,...,N$), the latent class is defined by a latent discrete variable denoted $c_i$ with value $g$ if subject $i$ belongs to latent class $g$ ($g=1,...,G$). Its distribution is described by the latent class membership probability $\pi_{ig}$ as follows:

\begin{equation}\label{regCI}
P({c_i=g})=\pi_{ig}=\dfrac{e^{\xi_{0g}+\bfs{X_{Ci}}^\top\xi_{1g}}}{\sum_{l=1}^{G} e^{\xi_{0l}+{X_{Ci}}^\top\xi_{1l}}}
\end{equation}
with constraints $\xi_{0G}=0$ and $\bf{\xi_{1G}=0}$ for reference class $G$. The probability $\pi_{ig}$ can be either defined as a regression on time-independent covariates $X_{Ci}$ or considered marginal by removing $X_{Ci}$ from the equation.  

\subsubsection{Class-specific repeated outcome distribution}
The latent class approach assumes that the outcomes under study have a class-specific distribution. In this work, we consider both quantitative repeated outcomes and time-to-event outcomes. Let $Y_{kij}$ denote the repeated value of outcome $k$ ($k=1,...,K$) for subject $i$ at occasion $j$ ($j=1,...n_{ki}$). The corresponding time of measurement is $t_{kij}$. Let $T^*_i$ denote the time to an event of interest with $L$ possible causes ($l=1,...,L$). The time-to-event can be right censored by censoring time $C_i$, and left truncated (as in our motivating application) with delayed entry $T_{0i}$. The observed time is denoted $T_i = \min(T^*_i,C_i)$ with indicator $d_i=l$ when the event of cause $l$ occurs before censoring and $d_i=0$ otherwise.

\paragraph{Quantitative repeated outcomes}

The class-specific trajectories of the repeated outcomes $Y_{kij}$ over time is modelled using a mixed model. The most common case is a continuous Gaussian outcome which trajectory over time is modeled by a linear mixed model specific to latent class $g$: 

\begin{equation}\label{LMM}
\left \{ \begin{array}{ll}
&Y_{kij} = Y^*_{ki}(t_{kij}) + \epsilon_{kij} \\
& \\
&Y^*_{ki}(t_{kij}) \mid_{c_{i}=g} ~=~ Z_{ki}(t_{kij})^T b_{ki}\mid_{c_{i}=g} + X_{Lki}(t_{kij})^{\top} \beta_{kg}  
\end{array}
\right.
\end{equation}

where $Y^*_{ki}(t_{kij})$ is the underlying level of the outcome without measurement error $\epsilon_{kij} \sim \mathcal{N}\left (0,\sigma_{\epsilon_{k}}^2 \right)$; $X_{Lki}(t_{kij})$ is the vector of covariates associated with the fixed effects $\beta_{kg}$, and $Z_{ki}(t_{kij})$ is the vector of covariates (most of the time limited to functions of time) associated with the individual random effects ${b}_{ki}$, with a class-specific distribution: ${b}_{ki} \mid_{c_{i}=g} = {b}_{kig} \sim \mathcal{N}\left (\mu_{kg},B_{kg} \right)$.

The mean trajectory over time $t$ in each latent class is thus $\text{E}(Y^*_{ki}(t)\mid_{c_{i}=g} )= Z_{ki}(t)^T \mu_{kg} + X_{Lki}(t)^{\top} \beta_{kg}$.

\paragraph{Quantitative repeated outcomes structured into latent dimensions}

When some markers measure the same underlying construct as systolic and diastolic blood pressure for instance, the markers can be structured into a reduced number $D$ of dimensions ($d=1,...D$), and the class-specific distribution applies to each dimension. 

Following previous works \cite{proust-lima2013}, we use a latent process approach to define the $D$ latent processes $\Lambda_{i}^d(t)$ from the repeated measures of the $K$ observed repeated outcomes. We consider that each outcome is a noisy measure of only one latent dimension using the following equation of observation: 

\begin{equation}\label{YLink}
H_k( Y_{kij}; \bfs{\eta_k} ) = \Lambda^{d(k)}_{i}(t_{kij}) + \epsilon_{kij} \\
\end{equation}

where $\Lambda^{d(k)}$ is the latent dimension measured by outcome $Y_k$, and $H_k$ is a bijective link function parameterized by $\bfs{\eta_k}$ which puts each outcome $k$ into the scale of the shared dimension $d(k)$.

Then, the class-specific trajectory of dimension $d$ is defined at any time $t$ ($t \in \mathbb{R}$) by a class-specific linear mixed model similar to the one given in Equation \eqref{LMM}: 

\begin{equation}\label{LambdaLMM}
\Lambda^{d}_{i}(t) \mid_{c_{i}=g} = Z_{di}(t)^T b_{di}\mid_{c_{i}=g} + X_{Ldi}(t)^{\top} \beta_{dg} 
\end{equation}

with $Z_{di}(t)$, $X_{Ldi}(t)$, $b_{di}$ and $\beta_{dg}$ having the same definition as in equation \eqref{LMM} except that they apply to the latent dimension $d$ instead of the outcome $k$. \\

Note that this more general formulation for the repeated outcomes modeling defined in Equations \eqref{YLink} and \eqref{LambdaLMM} includes different special cases: 

\begin{itemize}
\item[-]  one Gaussian continuous marker by dimension, that is Equation \eqref{LMM}, when $H_k$ is the identity and each marker has its own underlying dimension (i.e., $d(k)=k$ and $\Lambda^{d(k)}_{i}(t)=Y^*_{ki}(t)$). 

\item[-]  one non-Gaussian continuous marker by dimension when each marker has its own underlying dimension (i.e., $d(k)=k$) but $H_k$ is a nonlinear link function usually modelled using a basis of $M$ I-splines functions $(I_m)_{m=1,...,M}$, that is:
\begin{equation}\label{splines}
H_k(x;\bfs{\eta_k})=\eta_{0k} + \sum_{m=1}^M \eta_{mk}I_m(x) ~~~~ \text{with } x \in \text{range}(Y_k)
\end{equation}
In that case, some constraints need to be added so that $\Lambda^k$ has a determined dimension; this is usually done with a 0 mean in the reference category (for the location constraint), and either $\sigma_{\epsilon_{k}}^2 = 1$ or first diagonal element of $B_{1k}=1$ (for the dispersion constraint). 

\item[-]  multiple Gaussian and/or non-Gaussian continuous markers by dimension by appropriately define $H_k$ either as $H_k(x;\bfs{\eta_k})=\eta_{0k} + \eta_{1k}x$ (with $x \in \mathbb{R}$) for a Gaussian marker or according to equation \eqref{splines} for a non-Gaussian marker. As above, in this multivariate case, each $\Lambda^d$ needs to have a determined dimension with one constraint on the location, and one constraint on the dispersion. 

\end{itemize}

\subsubsection{Class-specific times-to-event distribution}

The time-to-event distribution in each latent class can be classically modelled within the cause-specific proportional hazard model framework where the class-specific instantaneous risk of event of cause $l$ is defined as follows:

\begin{equation}\label{CSevent}
\alpha_{il}(t) \mid_{c_{i}=g} =\alpha_{0lg}(t; \bfs{\zeta_g})\exp \left (\bfs{X_{Ti}}^\top \bfs{\delta_{lg}} \right )
\end{equation}

where $\alpha_{0lg}(t)$ is the instantaneous baseline hazard of cause $l$ in latent class $g$ and $\bfs{X_{Ti}}$ is a vector of covariates associated with fixed effects $\bfs{\delta_{lg}}$ (such fixed effects can also be considered as common over classes). 
Although any type of parametric hazards could be considered, we focus on Weibull hazards or approximate the baseline hazards by a small number of cubic M-splines. In addition, the baseline hazards can be either specific to each latent class or considered proportional across classes (i.e., $\alpha_{0lg}(t;\bfs{\zeta_g})=\alpha_0(t;\bfs{\zeta_0})\text{e}^{\zeta_{g}}$ with $\zeta_{G}=0$).

\subsection{Inference}

\subsubsection{Maximum Likelihood Estimation}

The joint latent class model defined in section \ref{JLCM} can be estimated in the maximum likelihood framework for a given number of latent classes $G$. Let $\bfs{\theta_G}$ denote the total vector of parameters for a $G$-class model. It includes all the parameters (subscripts $k$, $g$, $d$ are not reported here for readability) for: 
\begin{itemize}
\item[-] the latent class structure $\bfs{\xi}$;
\item[-]  the class-specific repeated outcomes distributions with fixed effects noted $\bfs{\beta}$ and $\bfs{\mu}$, variance-covariance of the random-effects $\text{vec}(\bfs{B})$ (parameterized using the Cholesky transformation), standard deviations of the errors $\bfs{\sigma_\epsilon}$ and parameters of the link functions when necessary $\bfs{\eta}$;
\item[-]  the time-to-event outcomes distribution $\bfs{\zeta}$ and $\bfs{\delta}$.
\end{itemize}

Thanks to the conditional independence between dimensions and time-to-event, the individual contribution $l_i$ to the likelihood based on the joint distribution of the repeated outcomes $\bfs{Y_i}=\{Y_{kij} \text{ with } k=1,...,K, j=1,...,n_{ik} \}$ and the time to event $(T_i,d_i)$ can be split as follows:  

\begin{equation} \label{loglikind}
\begin{array}{ll}
l_i(\bfs{\theta_G})& = f(\bfs{Y_i},(T_i,d_i);\bfs{\theta_G}) \\
& = \sum_{g=1}^{G} ~ P(c_i = g;\bfs{\theta_G}) \times  ~  f( \bfs{Y_i} | c_i=g;\bfs{\theta_G}) ~ \times ~ f((T_i,d_i) | c_i=g;\bfs{\theta_G}) \\
& = \sum_{g=1}^{G} ~\pi_{ig} ~\times ~ \prod_{d=1}^D f(\bfs{Y_{i}^d} \mid c_i = g;\bfs{\theta_G}) ~ \times ~ S_i(T_i \mid c_i=g,\bfs{\theta_G}) \prod_{l=1}^L  \alpha_{il}(T_i \mid c_i=g;\bfs{\theta_G})^{\bfs{1}_{d_i=l}} 
\end{array}
\end{equation}
 with $f$ the generic notation of a density function, and $P$ a probability function. 
 
The class-membership probability $\pi_{ig} $ is given in \eqref{regCI}, the instantaneous hazard $\alpha_i(T_i \mid c_i=g;\theta_G)$ is defined in \eqref{CSevent} and the corresponding survival $S_i(T_i \mid c_i=g,\theta_G) =  \exp \left ( - \sum_{l=1}^{L} \int_{0}^{T_i} \alpha_{il}(u \mid c_i=g;\bfs{\theta_G})du \right )$. Finally, the density function of the repeated outcomes is split into the product of the density functions of the subset of outcomes data, called $\bfs{Y_{i}^d}$, linked to each dimension $d$. Given the general formulation in equations \eqref{YLink} and \eqref{LambdaLMM}, the density function is:

\begin{equation}
 f(\bfs{Y_{i}^d} \mid c_i = g;\bfs{\theta_G}) = \phi(H(Y_i^d); m_i^d, V_i^d)\times \prod_{k=1}^{K(d)} \prod_{j=1}^{n_{ik}} J(H_k(Y_{kij}))
 \end{equation}
 where $\phi$ is the Gaussian density function with mean $m_i^d = X_{Ldi} \beta_{dg} + Z_{di} \mu_{dg}$ and variance $V_i^d = Z_{di}B_{dg}Z_{di}^{\top} + \Sigma_{di}$.
 $H(Y_i^d)$ denotes the vector $(H_k(Y_{ki}), k=1,\dots,K(d))^{\top}$ and $\Sigma_{di}$ is the diagonal matrix composed of values $\sigma_{\varepsilon_k}^2$ with $k = 1,\dots,K(d)$. $J(H_k(x))$ is the Jacobian of the link function $H_{k}$. See Proust-Lima et al. \cite{proust-lima2013} for further details. 
 
To take into account the delayed entry in $T_{0i}$ when times-to-event are left-truncated, the final individual contribution is divided by the probability to still be at risk of the events at entry:
\begin{equation}
l_i^{LT}(\bfs{\theta_G}) = \dfrac{l_i(\bfs{\theta_G})}{\sum_{g=1}^G \pi_{ig} S_i(T_{0i} \mid c_i=g,\bfs{\theta_G})}
\end{equation}

The final log-likelihood to be maximized is computed on all the subjects as $\mathcal{L} \left (\bfs{\theta_G} \right ) = \sum_{i=1}^N \log \left (l_i^{LT} \left (\bfs{\theta_G}\right ) \right)$. 

\subsubsection{Posterior Classification}

The posterior distribution of the latent classes can be derived from the observed data as: 
\begin{equation}
P \left ({c_i=g} \mid \bfs{Y_i},(T_i,d_i); \bfs{\theta}_G \right )= \dfrac{P \left (c_i=g; \bfs{\theta}_G  \right ) \times f \left (\bfs{Y_i} \mid c_i=g; \bfs{\theta}_G  \right ) \times f \left ((T_i,d_i)\mid c_i=g; \bfs{\theta}_G  \right ) }{f \left (\bfs{Y_i},(T_i,d_i);\bfs\theta_G \right )}
\end{equation}
where each element is calculated similarly as in the individual contribution to the log-likelihood in \eqref{loglikind}. 

This posterior distribution is classically computed at the point estimate $\bfs{\hat{\theta}_G}$, giving $\hat{\pi}_{ig} = P \left ({c_i=g} \mid \bfs{Y_i},(T_i,d_i); \bfs{\hat{\theta}_G} \right )$, and the posterior classification is derived: each individual is assigned to the latent class that provides the maximum individual posterior probability, that is the most likely class $\hat{c}_i = \argmax_{g=1,...,G} \hat{\pi}_{ig}$.

\subsubsection{Optimal number of latent classes}

Maximum likelihood is obtained for a fixed number of latent classes $G$, and the optimal number of latent classes thus needs to be \emph{a posteriori} determined. Selecting the optimal number of clusters in mixture problems is a wide area of statistical research \cite{celeux2018}. \\
Among information criteria, the Bayesian Information Criterion (BIC) is usually favored. Defined as $\text{BIC}(G) = -2 \mathcal{L}(\bfs{\hat{\theta}_G}) + p \log(N)$ with the lower the better, it was repeatedly shown to correctly select the optimal number of latent classes in different mixture situations \cite{nylund2007,morgan2015}. However, as it is a likelihood-based criterion, the BIC mainly focuses on the fit of the model to the data and may lead in some contexts to a poorly discriminant latent class structure \cite{celeux2018,bauer2003}. 

The discriminatory power of the latent class structure can be assessed by an entropy measure defined as $\text{EN}(G) = 1 + \dfrac{\sum_{i=1}^N \sum_{g=1}^G \hat{\pi}_{ig} \log(\hat{\pi}_{ig})}{N \log(G)}$ with values closer to one indicating higher discrimination of the classes \cite{morgan2016}. However, as built only from the posterior probabilities, this entropy measure completely neglects the fit of the model. 

When interested both in the fit and the clustering, the Integrated Classification Likelihood criterion (ICL) \cite{biernacki2000, morgan2016} has been considered. Defined as $\text{ICL}(G) = \text{BIC}(G) - 2 \times \sum_{i=1}^N \sum_{g=1}^G \mathbb{1}_{\hat{c}_i=g} \log(\hat{\pi}_{ig})$, this criterion penalizes the fit of the data by the discrimination power and thus can identify the latent class structure that provides the best balance between fit and discrimination. This is particularly useful in our exploratory context where we favor the identification of different relevant subphenotypes (i.e., classes) rather than the best fit to the data.   

\subsubsection{Multimodality}

One critical issue with latent class models is the multimodality of the likelihood and the potential convergence toward local suboptimal maxima \cite{hipp2006} with local optimizer. To ensure convergence toward the global likelihood maximum, we highly recommend the use of a gridsearch which replicates the estimation process for a large number of random initial values, and thus likely explores the entire parameter space and reaches the global likelihood maximum.

\subsubsection{Software}
The maximum likelihood estimation of the joint latent class model is implemented in lcmm R package \cite{proust-lima2017} with function mpjlcmm when considering $K \geq 1$ repeated markers and Jointlcmm when considering $K=1$ marker. Log-likelihood optimization is carried out by a robust Marquardt-Levenberg algorithm combined with stringent convergent criteria on the log-likelihood, the parameter and the first and second derivatives. This optimizer has been demonstrated to provide correct inference even in the case of complex log-likelihoods and/or relatively flat regions \cite{philipps2021}. 
\\
The package includes a gridsearch function for the parameter space exploration, and postfit functions for reporting the information criteria, posterior classification, goodness-of-fit, and predicted trajectories \cite{lcmmCran}. \\

\subsection{Association with external information}

After a latent class model estimation, one may want to assess the external predictors $\bfs{X_i}^\text{extern}$ of the latent class structure $c_i$, or one may want to determine how the latent class structure $c_i$ relates with an external outcome $Y_i^\text{extern}$ using regression techniques. In both cases, a naive approach consists in running the posterior regressions on an estimate of $c_i$ (for instance the most likely class $\hat{c}_i$) instead of the true unknown $c_i$, and neglecting the uncertainty in the estimate of $c_i$. The inference quality of this naive method, usually called "2-stage" \cite{elliott2020} or "3-step" method \cite{bakk2014}, depends on the discrimination of the latent classes. While it can provide negligible bias in case of highly separated classes (with high posterior probabilities, high entropy), it may become substantially biased in the case of rather poor discrimination  \cite{elliott2020, clark2009,bakk2014}. Alternatives consist either in corrections in the multi-step analysis to account for the uncertainty \cite{bakk2014}, or directly in the joint estimation of all the variables of interest including the external information $\bfs{X_i}^\text{extern}$ and/or $Y_i^\text{extern}$ to internally handle the measurement error in the latent class assignment \cite{elliott2020}. While the joint estimation naturally handles the latent nature of the classes and the uncertainty, one drawback is that the external information becomes part of the latent class model estimation and may slightly change the latent class structure. In many situations, the statistician wants to determine the latent class structure using a set of outcomes (in our case $Y_i$ and $(T_i,d_i)$), and relate this specific latent class structure with other external outcomes in posterior analyses (in our case $\bfs{X_i}^\text{extern}$ and/or $Y_i^\text{extern}$).  

To both separate the estimation process of the latent class structure from the posterior analyses (as done in multi-step techniques), and account for the latent nature of the class structure (as done in the joint estimation technique), we used an intermediate approach. We considered the joint likelihood including $Y_i$, $(T_i,d_i)$ and the external information, either $\bfs{X_i}^\text{extern}$ or $Y_i^\text{extern}$. However, we did not re-estimate all the parameters involved, only those related to $\bfs{X_i}^\text{extern}$ or $Y_i^\text{extern}$ as explained below.

\paragraph{Case 1. The latent class structure as a predictor of an external outcome $Y_i^\text{extern}$} 

We consider the case of a continuous marker $Y_i^\text{extern}$ either measured once or repeatedly over time. The same generic model as defined in \eqref{YLink} and \eqref{LambdaLMM} can be applied to $Y_i^\text{extern}$ with its specific underlying level $\Lambda^{D+1}$. Note that the use of the identity in \eqref{YLink} and no random effects in \eqref{LambdaLMM} provides for instance a linear model for cross-sectional data. Let's define $\bfs{\psi_G^\text{extern}}$ the total vector of parameters for this external outcome. By including the external outcome, the contribution to the joint log-likelihood previously defined in \eqref{loglikind} now includes a fourth element: 

\begin{equation} \label{loglikind_extern}
\begin{array}{ll}
l_i^{Y^\text{extern}} \left (\bfs{\theta_G},\bfs{\psi_G^\text{extern}} \right)& = f \left (\bfs{Y_i},(T_i,d_i),Y_i^\text{extern};\bfs{\theta_G}, \bfs{\psi_G^\text{extern}} \right ) \\
& = \sum_{g=1}^{G} ~ P(c_i = g;\bfs{\theta_G}) \times   f( \bfs{Y_i} | c_i=g;\bfs{\theta_G}) \times  f((T_i,d_i) | c_i=g;\bfs{\theta_G}) \times \\
& ~~~~~~~~ f(\bfs{Y_{i}^\text{extern}} \mid c_i = g;\bfs{\psi_G^\text{extern}}) 
\end{array}
\end{equation}

The posterior regression for $Y_i^\text{extern}$ can be estimated by maximising $\mathcal{L}^{Y^\text{extern}}  \left ( \widehat{\bfs{\theta}}_G,\bfs{\psi_G^\text{extern}} \right ) = \sum_{i=1}^N \log \left ( l_i^{Y^\text{extern}}  \left (\widehat{\bfs{\theta}}_G,\bfs{\psi_G^\text{extern}} \right) \right )$ according to $\bfs{\psi_G^\text{extern}}$. 

\paragraph{Case 2. External information $\bfs{X_i}^\text{extern}$ as the predictor of the latent class structure}

This case is usually sought when the joint latent class model does not already include predictors of the latent class structure in equation \eqref{regCI}. External predictors $\bfs{X_i}^\text{extern}$ can be easily included a posteriori by updating equation \eqref{regCI} with $\bfs{X_{Ci}}=\bfs{X_i}^\text{extern}$.
The estimation technique is then very similar as for case 1. We define $\bfs{\xi_G^\text{extern}}$ the total vector of parameters involved in the updated formula \eqref{regCI} according to $\bfs{X_i}^\text{extern}$, and we consider the following contribution to the joint log-likelihood where the component involving parameters $\bfs{\xi_G^\text{extern}}$ is now re-estimated. Note that for clarity we mention here the condition on $\bfs{X_i}^\text{extern}$: 

\begin{equation} \label{loglikind_extern}
\begin{array}{ll}
l_i^{X^\text{extern}} \left (\bfs{\theta_G},\bfs{\xi_G^\text{extern}} \right)& = f \left (\bfs{Y_i},(T_i,d_i)| \bfs{X_i}^\text{extern};\bfs{\theta_G}, \bfs{\xi_G^\text{extern}} \right ) \\
& = \sum_{g=1}^{G} ~ P(c_i = g| \bfs{X_i}^\text{extern};\bfs{\xi_G^\text{extern}}) \times   f( \bfs{Y_i} | c_i=g;\bfs{\theta_G}) \times  f((T_i,d_i) | c_i=g;\bfs{\theta_G}) 
\end{array}
\end{equation}

The posterior regression of $c_i$ according to $\bfs{X_i}^\text{extern}$ can be estimated by maximising $\mathcal{L}^{X^\text{extern}}  \left ( \widehat{\bfs{\theta}}_G,\bfs{\xi_G^\text{extern}} \right ) = \sum_{i=1}^N \log \left ( l_i^{X^\text{extern}}  \left (\widehat{\bfs{\theta}}_G,\bfs{\xi_G^\text{extern}} \right) \right )$ according to $\bfs{\xi_G^\text{extern}}$.

\section{Simulation Study}

\rev{We carried out a simulation study to explore the finite sample properties of the estimation procedure of the multivariate joint latent class model. The simulation study is fully detailed in Supplementary Materials, Section 1. This includes the generation algorithm, the description of the different scenarios, the results and their interpretation, as well as a replication script in Section 3.
Briefly, we generated series of 300 samples of 250, 500 or 750 subjects constituted of 3 latent classes, $K=$2 or 3 repeated outcomes with class-specific linear trajectories and one or two competing-cause survival times under class-specific Weibull risks or proportional piecewise exponential hazards with parameters chosen to achieve different levels of entropy (between 63\% and 83\%) and different proportions of events (between 18\% and 77\%). This lead to 9 scenarios (see description in Table S1), 6 of them run for the three sample sizes. For each sample, the model was estimated using a grid of 100 random sets of initial values.}

\begin{figure}[!p]
\centering\includegraphics[width=0.85\textwidth]{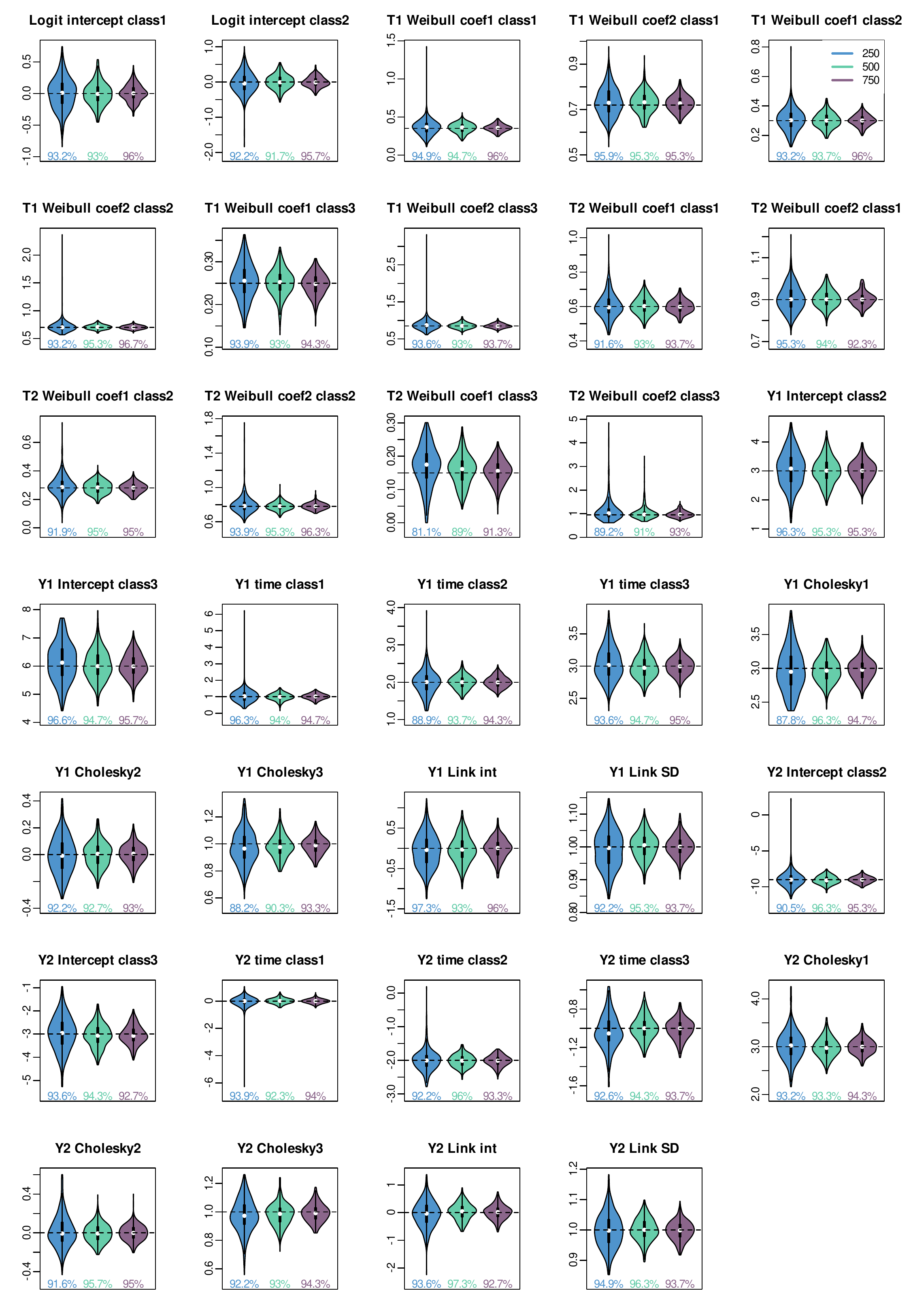}
\caption{Violin plots of the parameters estimates in the 300 replicates of simulation Scenario 4 for 250, 500 and 750 subjects. The specified model included 3 latent classes, 2 markers with linear trajectory (Y1, Y2) and a class-specific an cause-specific Weibull risk of event for two causes of event (T1, T2).}
\label{simviolin}
\end{figure}

\rev{All the simulation results are reported in supplementary Tables S2 to S25. We also report the estimated parameters along with coverage rates of the 95\% confidence interval in Figure \ref{simviolin} for scenario 4 that included two competing causes of event and an entropy of 0.71. Overall, the simulation results illustrate the correct estimation of the parameters in all the scenarios with negligible bias and good coverage rates of the 95\% confidence intervals for samples of 500 and 750 subjects. For samples of 250 subjects, the estimates were generally well estimated although scenarios 3, 4, 5 revealed a small bias and too low coverage rate for the class-specific survival parameters in cases where the number of events in the class was very low (<10 events). This was explained by rare extreme values and these estimations remained overall good over the replicates despite the very low proportion of events (see violin plots in Figure \ref{simviolin} and Figure S3, for scenarios 4 and 5, respectively).}

\rev{Scenario 8 also aimed to illustrate how a three-dimensional model could help identify further heterogeneity compared to uni-dimensional models (with 3 classes identified while uni-dimensional models could only retrieve 2 classes) (Supplementary section 1.3.8., Tables S23, S24).}

\section{Application to MSA progression} \label{appli}

\begin{figure}[!p]
\centering\includegraphics[width=0.9\textwidth]{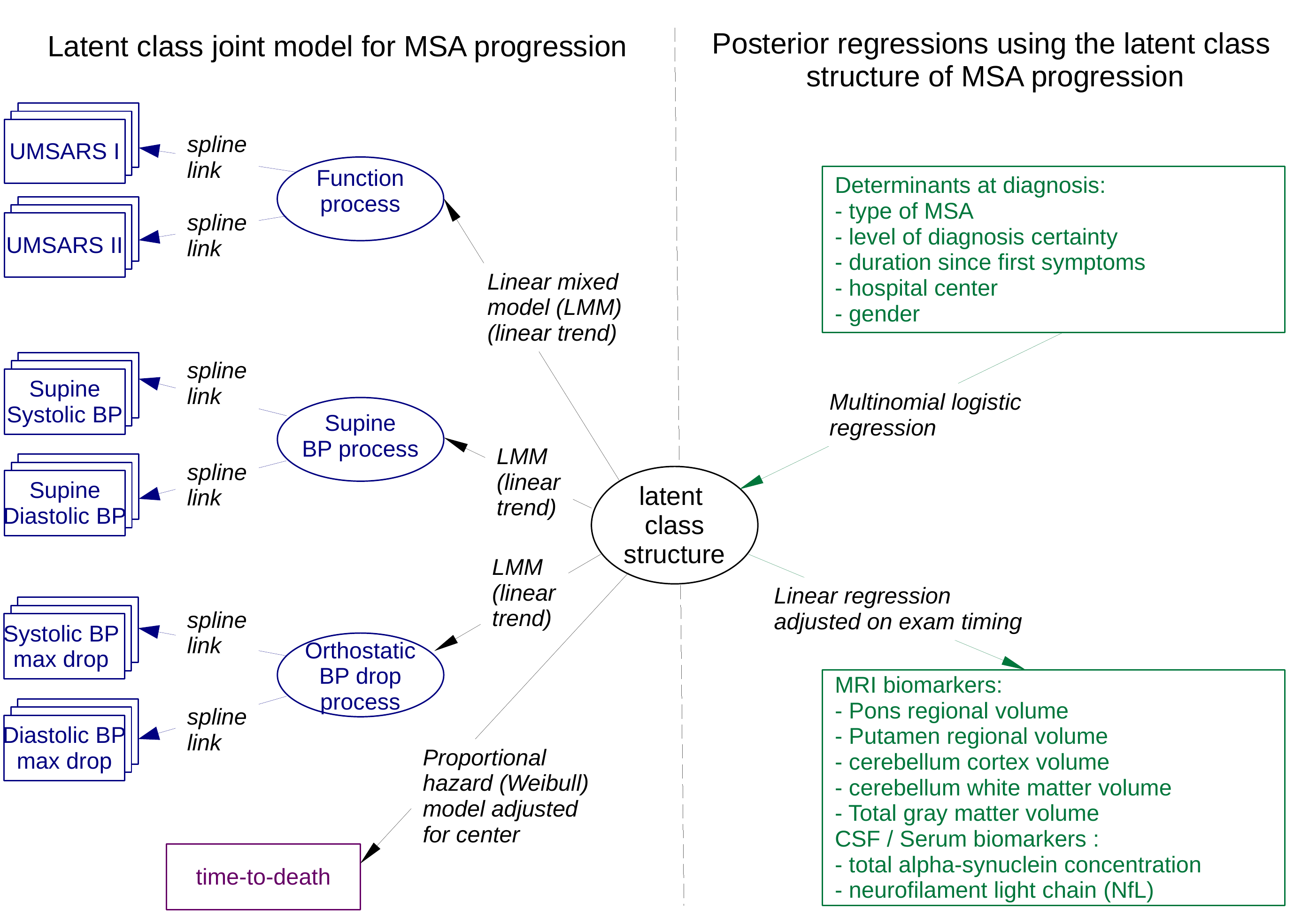}
\caption{Diagram summarizing the definition of the joint latent class model for MSA progression (left part), and the associated posterior analyses (right part). Are reported the chosen specifications for each submodel (\rev{carefully} determined in preliminary separated analyses \rev{after comparison with alternative candidates} according to AIC).}
\label{fig_schema}
\end{figure}

We applied the joint latent class methodology to describe the progression of the 6 markers measured repeatedly over time and grouped into three dimensions: \textbf{function} with the sumscores UMSARS I and II, \textbf{supine blood pressure} with the diastolic and systolic measures, \textbf{orthostatic BP drop} with the maximum decreases in systolic and in diastolic blood pressure, as long as \textbf{time to death}. The specification of the model is summarized in Figure \ref{fig_schema}. \rev{Each parametric assumption was carefully assessed in separate preliminary analyses for $G=1$ by comparing different candidates according to the Akaike Information Criterion (AIC) and BIC}. For instance, better AIC were found when considering a Weibull baseline risk function compared to a basis of M-splines with 3 internal knots, \rev{when considering quadratic splines for the link functions compared to linear transformations,} or when considering a linear trajectory of the dimensions over time compared to nonlinear trajectories approximated by natural cubic splines or polynomials. \rev{In addition, although the three underlying dimensions were defined based on clinical knowledge, we compared a latent process model (where the two constituting markers are assumed to measure the same underlying process) with a bivariate mixed model (where each marker has its own trajectory, and random effects are correlated between markers). The assumption of an underlying process was reasonable for the three dimensions under $G=1$. The AIC and BIC concluded to the selection of the latent process model for Supine BP (AIC/BIC = 26170.01/26275.45 and 26175.04/26245.33 for the bivariate model and for the latent process model, respectively) and Orthostatic BP (AIC/BIC = 25483.22/25588.67 and 25471.33/25541.63 for the bivariate model and for the latent process model, respectively). For the Function process, the bivariate mixed model provided a better fit (AIC/BIC = 21731.81/21837.25 and 21909.83/21980.13 for the bivariate model and for the latent process model, respectively) but the latent process model remained reasonable and was clinically justified as both UMSARS-I and UMSARS-II scores measure Function degradation. }

\subsection{Selection of the number of latent classes}
\begin{figure}[!p]
\centering\includegraphics[width=0.7\textwidth]{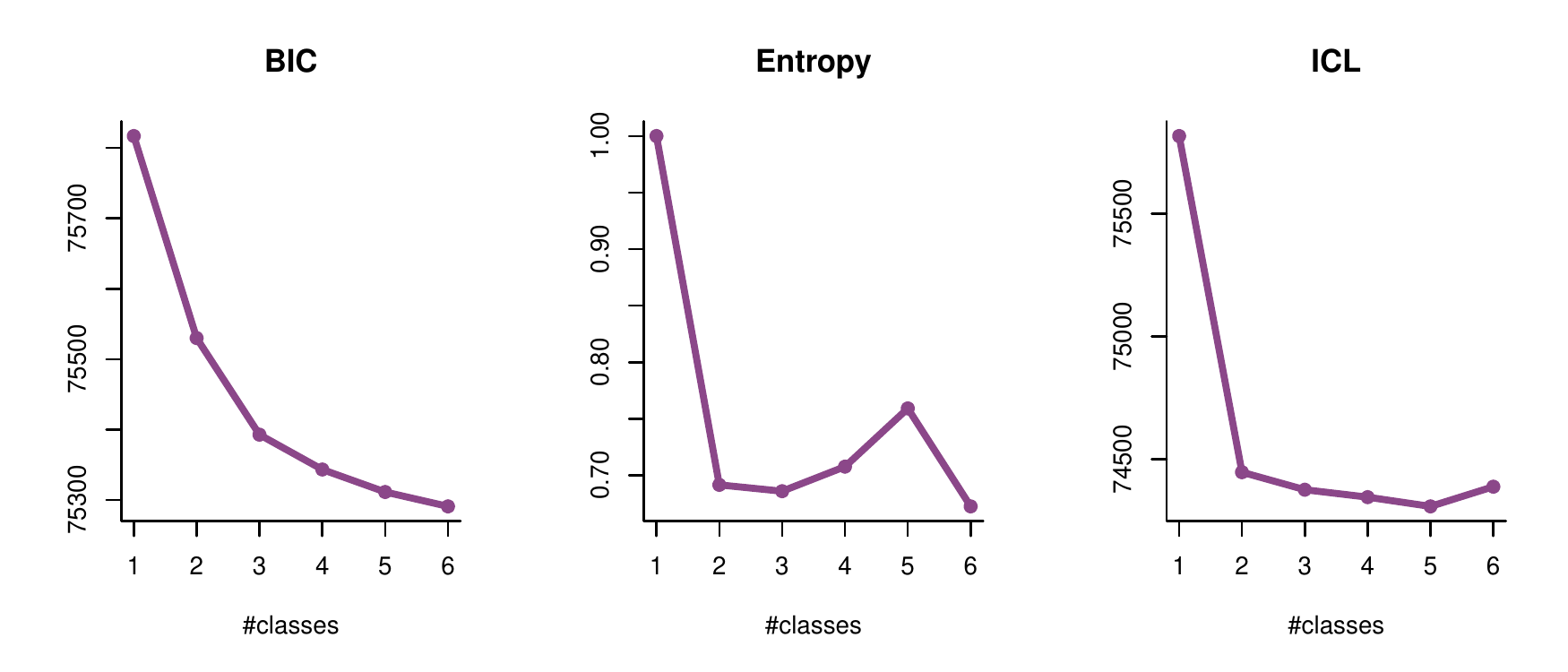}
\caption{Comparison of models considering from 1 to 6 latent classes. Are reported the BIC for goodness-of-fit assessment, Entropy for discriminatory assessment, and ICL for an overall assessment.}
\label{summaryplot}
\end{figure}

Joint latent class models assuming between 1 and 6 latent classes were repeatedly estimated using a grid of 100 random initial values. Figure \ref{summaryplot} summarizes the three statistical criteria (BIC, Entropy, ICL) used for determining the optimal number of latent classes. While the goodness-of-fit was gradually improved when adding a new latent class, the entropy was clearly better for the 5-class model (Entropy at 0.76) suggesting that although the 6-class was even closer to the data, it did not provide a sufficiently high separation of the patients. As the ICL which accounts for both goodness-of-fit and discrimination also favored the 5-class model, we retained these 5 subphenotypes of MSA clinical progression. The Sankey plot (displayed in Figure S5 in supplementary materials) describes the sequence of latent class splits with the increasing number of classes. 

\subsection{5 subphenotypes of MSA progression}

\begin{figure}[!p]
\centering\includegraphics[width=0.98\textwidth]{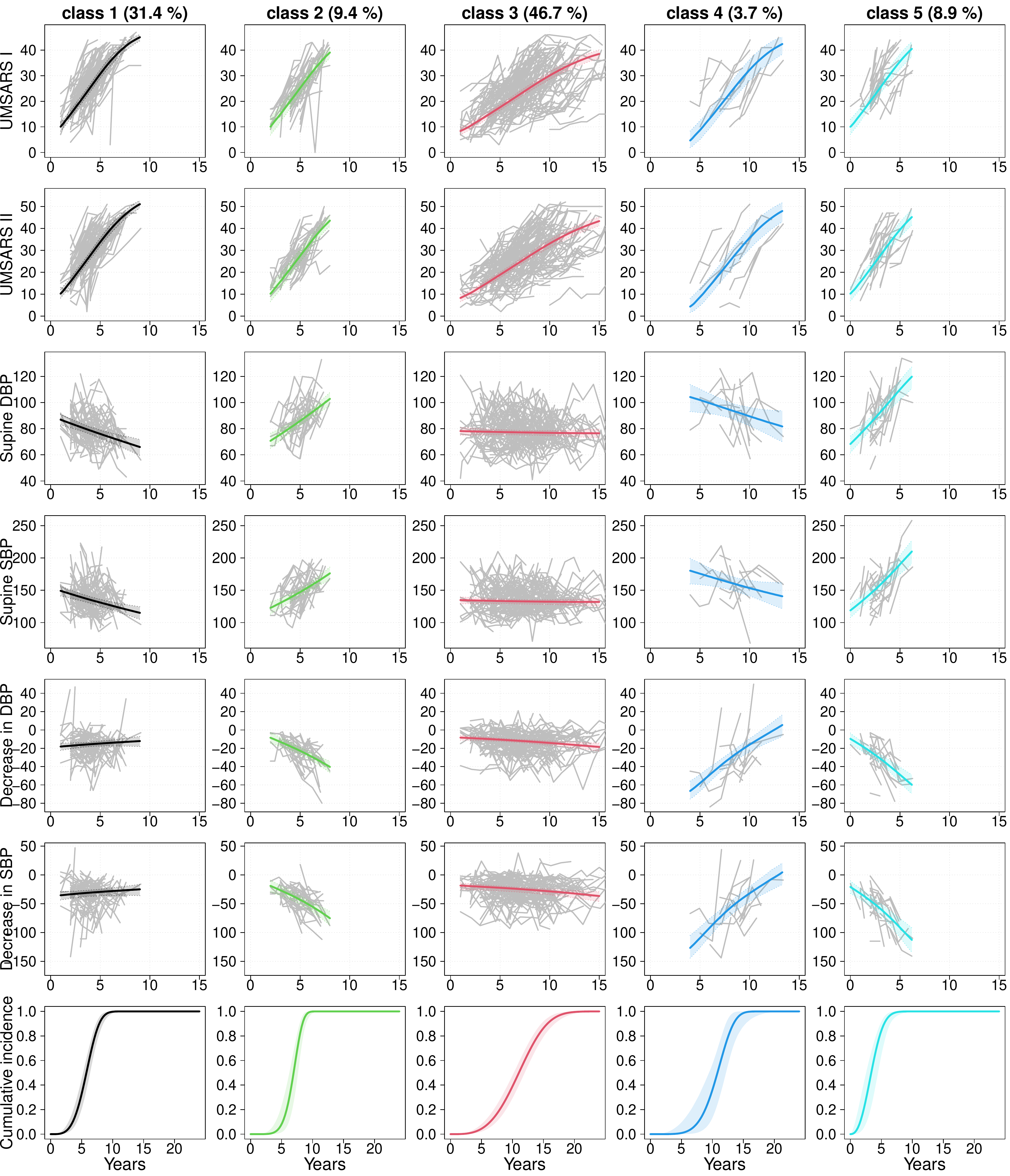}
\caption{Predicted trajectories of the markers (and 95\% in shades) and predicted death probability in the 5 latent classes. Are also reported in grey lines the observed trajectories of the patients \emph{a posteriori} classified into the latent class.}
\label{PredClass}
\end{figure}

The mean trajectories of the 6 markers and the predicted death probability characterizing the 5 subphenotypes of MSA progression are reported in Figure \ref{PredClass}. The 5 subphenotypes differed by the shape and speed of progression of the three dimensions, and the risk of death. The largest class (Class 3) with 46.7\% of the sample was characterized by a much slower deterioration of the function (UMSARS I and II) than others, and a relatively stable level of supine BP, and slight increase in orthostatic BP drop. The second largest class (Class 1) with 31.4\% of the sample was characterized by a fast deterioration of the function but also a decrease in supine BP over time and rather stable or slight decrease in orthostatic BP drop. The classes 2 and 5 comprised around 9\% of the sample each and were both characterized by similar shapes of clinical progression: fast deterioration of the function, increase in supine BP and aggravation of the orthostatic BP drop. However, the timing was different. The patients from class 5 had a progression beginning right after the first symptoms while this progression was slightly delayed in class 2. Finally, the smallest class (Class 4), which included 3.7\% of the sample, was characterized by a fast deterioration of the function, and at the same time, a decrease in supine BP and orthostatic BP drop which makes it very peculiar. As shown in the Sankey plot (Figure S4 in supplementary materials), the smallest latent class 4 was only identified when considering a fifth class. As this small class substantially differs from the others, this probably explains the gain in entropy observed between the 4- and 5-class model.

The risk of death was substantial in all the classes with a probability of death reaching 1 in all classes by 15 years after the first symptoms. It followed the functional degradation with a more progressive risk of death in class 3 compared to others and earlier risk for class 5 and 1 compared to classes 3 and 4. 

\subsection{Determinants of the latent classes}

In posterior multinomial logistic regression, we assessed the determinants of the 5 subphenotypes (Figure S6 in supplementary materials). No difference was observed according to sex or MSA center. However, as previously identified, the duration between the first symptoms and the diagnosis substantially differed according to the latent class with later diagnoses for latent classes 3 and 4, and earlier diagnoses for latent classes 1 and 5. Patients with a cerebellar presentation of the disease were more likely classified in latent class 4 compared to others, and  less likely classified in the fast progressors of class 1. Finally, patients with a probable diagnosis were much more likely classified in any other class than class 3 compared to patients with a possible diagnosis. 

\subsection{Association with MRI and fluid biomarkers}

Understanding the underlying biological mechanisms of MSA is particularly crucial for therapeutic development. Indeed, beyond MSA patients care, MSA constitutes a fast model for the group of $\alpha$-synucleinopathies including Parkinson's disease. As such, identifying potential therapeutic targets, or differential biological mechanisms is of high importance. The classification may be useful to explore how new biomarkers differ according to this parcimonious summary of the MSA clinical progression. 

\begin{figure}[!p]
\centering\includegraphics[width=0.9\textwidth]{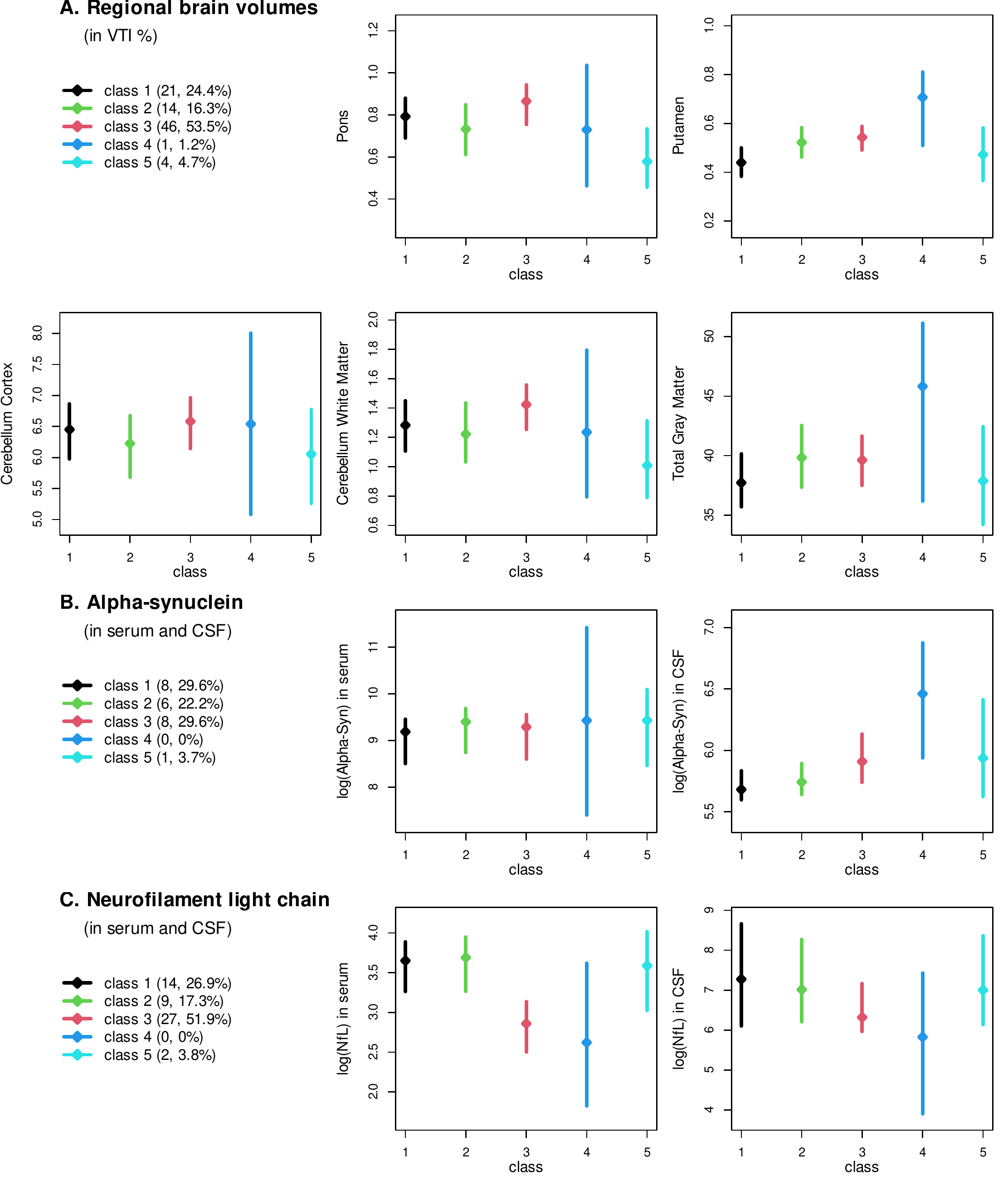}
\caption{Posterior MRI and fluid biomarkers differences across classes predicted in separated linear regressions run on subsamples of 86 (A), 23 (B) and 52 (C) patients. Regressions are adjusted for the exam timing, and account for the uncertainty in the latent class assignments.}
\label{postreg}
\end{figure}

We focused here on MRI biomarkers with 5 brain regions of interest (N=86 patients), and serum and CSF measures of total $\alpha$-synuclein (N=23) and of NfL (N=52). Due to the small sample sizes, we focused mainly on the differences between the two largest classes, class 1 of fast progressors and class 3 of slow progressors. The posterior linear regressions adjusted for the exam timing displayed in Figure \ref{postreg} suggested a more preserved MRI structure for class 3 than class 1 with in particular a larger putamen volume and total gray matter. The concentration of NfL, a marker linked to the aggressiveness of axonal injury, also tended to be higher in the fast progression class 1 (classes 2 and 5 too) compared to the slow progression class 3, especially in the serum, while the CSF concentration of total $\alpha$-synuclein was slightly lower for the fast progressors. 

\subsection{Goodness-of-fit assessment}

We followed the strategy described in Proust-Lima et al. \cite{proust-lima2016,proust-lima2014} to assess the goodness of fit of each part of the final joint latent class model:

\begin{itemize}
\item longitudinal data: we compared the trajectories of weighted mean predicted values in each class to the trajectories of weighted mean observations. Specifically, observation times were split into intervals. Then subject-and-class-specific marker predictions computed at each observation point of an interval were averaged with weights corresponding to the posterior individual probability to belong to each class. The same strategy was used for the observations. Applied to the selected 5-class model, it showed that the weighted averaged predictions of each marker were very close to the weighted averaged observations (Figure S7 in supplementary materials).

\item survival data: because of the delayed entry, we did not compare the predicted and observed survival functions or cumulative hazards in each class. Instead we compared the weighted mean of class-and-subject-specific predicted instantaneous risk functions to the instantaneous risk function in each class estimated by a weighted piecewise hazard model with knots every two years. 95\% confidence interval was obtained by non-parametric bootstrap with 200 samples. Applied to the selected 5-class model, it showed that the weighted predicted risks of death were close to the observed ones (Figure S8 in supplementary materials). 

\item classification: the quality of the classification obtained from the 5-class joint model was assessed by the
posterior classification table (Table S30 in Supplementary Materials). In each class, the mean posterior
probability of belonging to this class ranged from 77.7\% in Class 1 to 88.6\% for the slower progression class 3, and 89.4\% for the small and peculiar class 4, indicating a clear discrimination between the latent classes. 
\end{itemize}

\subsection{\rev{Comparison with unidimensional joint latent class models}}

\rev{In secondary analyses, we estimated a joint latent class model on each dimension taken separately. The model specification and the strategy of analysis remained the same. Four classes were identified for the function (12.5\%, 29.8\%, 41.8\%, 15.9\% - with a gradient of increasing slopes) and two small classes of rapid changes were distinguished for the supine BP (5.4\% - with a substantial increase) and orthostatic BP (11.7\% - with an amplification of the orthostatic hypotension), the rest of the sample having a rather stable progression of both BPs (Figure S9). When comparing the posterior classifications with the one of the 5-class multidimensional joint model, the four latent classes of Function progression discriminated mainly classes 1, 3 and 4 while the two small classes of Supine and Orthostatic BP rather corresponded to latent classes 2 and 5. }

\section{Discussion} \label{discussion}

Given the complexity of some diseases and the richness of the data collected in cohort studies, methods to summarize multivariate longitudinal information, and capture heterogeneity become real assets in biostatistics. With this work and the associated implementation in the R package lcmm (function mpjlcmm), we provide a relevant and effective solution \rev{validated in simulations} for summarizing information from multivariate markers measured repeatedly over time and clinical endpoints. In the MSA example, the approach summarized 6 marker trajectories and risk of death into 5 subgroups of patients with different profiles of progression that suggest distinct subphenotypes of the disease. 

In addition to unraveling a heterogeneous clinical progression, this method provides a simple parcimonious summary of complex disease progression that can then be used to explore new research directions, and markers of interest. For instance, in MSA, although based on a very small subset of patients, posterior analyses of the classification suggested a preserved MRI structure in the slow progression class 3 compared to other classes, and higher NfL for the rapid progression classes confirming the higher aggressiveness of these profiles. A slightly lower concentration in CSF total $\alpha$-synuclein was also observed for the fast progressors which suggests a higher pathological sequestration of $\alpha$-synuclein in the brain for these patients \cite{mollenhauer2011}. The differences in total $\alpha$-synuclein were small across classes compared to the differences in NfL. This is probably due to the fact that $\alpha$-synuclein is a marker of the pathophysiological process rather than a marker of progression. Although further research is needed to confirm these observations, they illustrate how this statistical methodology opens up perspectives in a complex disease such as MSA to improve the understanding of pathophysiological processes. \\

The statistical model relies on the assumption of conditional independence between processes, i.e. all the correlations between such rich data can be captured by a few latent classes. We are aware that this assumption is likely violated. However, our objective was not to properly assess the nature of the association between the markers and clinical endpoints but to explore and identify heterogeneous profiles of progression that could be used as a parcimonious summary to be considered for external analyses. In this context, we did not further test the independence assumption. We relied instead on criteria such as the entropy or the ICL (which gives a balance between discrimination ability and goodness-of-fit) to assess the quality of the classification, and its discrimination ability. With a final entropy of 0.76, and mean posterior class-membership probabilities between 0.78 and 0.89, the 5 latent classes showed a very convincing split of the population into distinct profiles of progression that can be referred to as subphenotypes. 

\rev{We carried out additional simulations to explore the behavior of the method under a residual correlation of 0.2 and 0.3 between the markers (See subsection 1.3.9 in supplementary materials). Overall, the parameters’ distributions under misspecification showed small bias but were not too much impacted (Figure S4, Tables S25-S27). Moreover, the percentage of individuals correctly classified remained similar to the one under true conditional independence (about 85\% overall).}

Previous works had formally addressed the issue of the conditional independence in joint latent class models by focusing on simpler settings (single repeated marker and event time). Jacqmin-Gadda et al. \cite{jacqmin-gadda2010} and Proust-Lima et al. \cite{proust-lima2017} developed score tests to evaluate whether there is residual dependence between a repeated marker and clinical endpoints. The same strategy could be undertaken to provide a score test for the conditional independence adapted to the multivariate context. Andrinopoulou et al. \cite{andrinopoulou2020} and Liu et al. \cite{liu2015} also developed joint models that included both marker-event dependence on latent classes and on shared random-effects. However, these hybrid models already showed substantial numerical problems in the univariate situation, and their applications were limited to a too small number of classes to be useful in our context of progression summary. We thus leave these directions of development for future research. \\

Beyond the multivariate nature of the processes in play, the study of chronic disease progression usually induces additional complexities that our methodology and the associated software can handle: (i) Gaussian and non-Gaussian distributions of markers managed by defining parameterized link functions following previous work of the authors \cite{proust-lima2013}, (ii) markers measuring the same underlying dimension handled by shared latent processes, (iii) delayed entry taken into account in the estimation procedure, (iv) competing risk setting with cause-and-class specific proportional hazard models (although not detailed here as not relevant in the MSA context, it is included in the software solution). Still, some issues are left for future improvements. First, although the theory could apply to other natures of repeated markers, especially within the exponential family with generalized linear mixed models applied to the latent dimensions, we only focused on continuous markers. Second, we described the trajectory according to the time since the first symptoms under the assumption that, at their inclusion in the cohort, the patients were able to accurately determine the time since their first symptoms. Dealing with this type of uncertainty calls for methodologies based on latent disease time \cite{li2017} that could be combined with the latent class approach in the future. 
\rev{It is important to recall that our solution, although flexible, remains fully parametric. As such, each part of the model (e.g., existence of underlying latent dimensions, link functions, shape of trajectory, baseline risk functions, selection of the number of latent classes) has to be carefully determined in preliminary analyses and posterior evaluations. In the application, we postulated notably the existence of underlying processes, each one measured by two markers. This was clinically justified and seemed reasonable given the data both in preliminary analyses and in the posterior comparison of predictions versus observations (Figure S7). Another essential caution with the use of joint latent class models and mixture models in general is that they constitute flexible approaches to model asymmetric distributions or heavy tailed distributions even in the absence of a real latent class structure (see for instance Bauer and Curran (2003)\cite{bauer2003} and discussants). This is why in this work where the latent class structure was central, we did not rely only on the goodness-of-fit but also on the discrimination performances with the entropy (Figure \ref{summaryplot}) and posterior class-membership probabilities (Table S30).} \\

Linking latent classes to external outcomes, as done in MSA with MRI, CSF and plasma markers, constitutes one direction of research of its own due to the difficulty to account for the uncertainty in the estimated latent class structure \cite{clark2009,bakk2014, bakk2018,elliott2020}. In our work, we chose to directly integrate the external outcomes into the joint model program to correctly handle the uncertainty on the latent class membership, as suggested by others in a different latent class framework \cite{elliott2020,bakk2018}. \\

In conclusion, the \rev{multi-dimensional} latent class methodology described here is a powerful, flexible and effective tool for exploring disease progression especially in complex settings as encountered in MSA with different markers of different dimensions and no clear biological assumption behind. It opens to a deeper understanding of the disease progression, and exploration for phenotypes differences. Although limited in our motivating example to several MRI, CSF and plasma markers, posterior analyses based on latent classes can also apply in high dimensional contexts with omics information for instance.


\section*{Acknowledgments}
Computer time was provided by the computing facilities MCIA (Mésocentre de Calcul Intensif Aquitain) at the University of Bordeaux and the University of Pau and Pays de l’Adour. We thank the INSERM/UPS UMR1214 Technical platform for the MRI acquisition and Centre d'Investigation Clinique (CIC) for the global coordination of the clinical study within INSERM-DGOS 2013-2014 grant.

\subsection*{Funding}
This work was funded by the French National Research Agency (grant number ANR-18-CE36-0004 for project DyMES), by Appel d'Offre Interne of Bordeaux university hospital (grant AOI 2011 for project BIOMAS, grant AOI 2014 for project COGAMS), by PHRC  in 2012 (DGOS 161), association PSP France for project Biopark and association ARAMISE. MRI data acquisition was supported by a "Recherche clinique translationnelle" grant from INSERM-DGOS 2013-2014.

\subsection*{Author contributions}
Cécile Proust-Lima (CPL) and Viviane Philipps (VP) developed the statistical methodology and carried out the validation by simulations. Tiphaine Saulnier (TS), CPL and Alexandra Foubert-Samier (AFS) conceived, designed and carried out the data analysis. AFS, Wassilios Meissner (WM), Anne Pavy-Le Traon (APLT), Patrice Péran (PP) and Olivier Rascol (OR) provided the data. CPL drafted the manuscript. VP, TS, AFS substantially revised the manuscript. All authors read and approved the final manuscript. All authors take responsibility for the integrity of the data and the accuracy of the data analysis. 

\subsection*{Financial disclosure}
None reported.

\subsection*{Data Availability}
Programs to simulate data and validate the inference method are available in the supplementary material of this article. 
MSA data are not shared.

\subsection*{Conflict of interest}
The authors declare no potential conflict of interests.

\section*{Supporting information}
The following supporting information is available as supplementary materials: additional figures and tables for the application (Section 1) and a simulation study with R scripts for replication (Section 2).

\bibliography{mpjlcmm_HAL_Arxiv_V2}%

\end{document}